\documentclass[aps,prl,reprint]{revtex4-1}

\usepackage[utf8]{inputenc}
\usepackage[T1]{fontenc}
\usepackage{hyperref}
\usepackage{graphicx}
\usepackage{etoolbox}
\usepackage[lcgreekalpha]{stix}
\usepackage{mathtools}
\usepackage{siunitx}
\usepackage{microtype}

\renewcommand{\section}[1]{\emph{#1.}---}

\renewcommand{\vec}{\mathbfit}

\newcommand{\as}{a_\mathrm{s}}

\newcommand{\Vs}{V_\textrm{s}}
\newcommand{\uppi}{\mathrm{\pi}}

\graphicspath{{figures/}}

\newtoggle{includesupplementary}
\toggletrue{includesupplementary}

\begin{document}
\title{Polyatomic trilobite Rydberg molecules in a dense random gas}

\author{Perttu J. J. Luukko}
\email{perttu.luukko@iki.fi}
\altaffiliation[Current address: ]{Tampere University of Technology, Laboratory of Physics, Tampere, Finland}
\author{Jan-Michael Rost}
\affiliation{Max Planck Institute for the Physics of Complex Systems, Dresden, Germany}

\begin{abstract}

Trilobites are exotic giant dimers with enormous dipole moments.
They consist of a Rydberg atom and a distant ground-state atom bound together
by short-range electron-neutral attraction.
We show that highly polar, polyatomic trilobite states unexpectedly persist and
thrive in a dense ultracold gas of randomly positioned atoms.
This is caused by perturbation-induced quantum scarring and the localization of
electron density on randomly occurring atom clusters.
At certain densities these states also mix with a s-state, overcoming selection
rules that hinder the photoassociation of ordinary trilobites.

\end{abstract}

\date{\today}

\makeatletter
\hypersetup{pdfauthor={Perttu J. J. Luukko, Jan-Michael Rost},pdftitle={\@title}}
\makeatother

\maketitle


If an atom is excited to a Rydberg state~$n$ in a sufficiently dense gas, the
range of its electron cloud ($\propto n^2$) can be large enough to reach a
nearby ground-state atom. For some elements the short-range electron-atom polarization
interaction is attractive, and in very low
temperatures it can create a bound state -- an exotic giant molecule thousands
of atomic units in size~\cite{trilobite}.

Rydberg molecules separate into two classes due to quantum defects affecting
low angular momentum states of the Rydberg electron. Low-$l$ molecules are
created from energetically separated states with~$l\lesssim 2$ (typically an
s-state). These molecules have relatively uniform electron densities and
shallow Born--Oppenheimer (BO) potentials. In contrast, quasidegenerate
high-$l$ states combine into much more deeply bound molecules where the Rydberg
electron localizes on the distant ground-state atom. As a result, these
so-called trilobite molecules have permanent dipole moments in the kilodebye
range, making them very sensitive to external fields and a promising candidate
for realizing exotic, strongly correlated systems.

Both classes of Rydberg molecules have been observed
experimentally~\cite{greene06, bendkowsky2009, li, tallant, bellos, krupp,
anderson2, booth, kleinbach}. Because selection rules inhibit the
photoassociation of pure high-$l$ trilobites, the production of highly polar
Rydberg molecules requires circumstances where the two classes mix into a
hybrid with a suitable mixture of low- and high-$l$ character. This has been
only recently achieved for certain elements and Rydberg states using nearly
integer quantum defects~\cite{booth,eilesgreene} or resonant hyperfine
splitting~\cite{kleinbach}.


In sufficiently high densities the Rydberg electron can reach and bind several
ground-state atoms. Polyatomic s-state Rydberg molecules with up to five
constituent atoms have been observed in experiments~\cite{bendkowsky2010,gaj}.
For an isolated s-state the effect of additional atoms to the electronic wave
function can be neglected and very high densities can be studied using a
mean-field model that adds up contributions for each atom from the dimer BO
potential~\cite{schlagmuller,liebisch}. This approach does not apply for the
trilobite class, where the electron density depends on the positions of all
perturbing atoms.

Polyatomic trilobites have been studied in constrained geometries
where all atoms share a common distance~$R$ to the Rydberg
core~\cite{LR06,eiles}. This allows solving the electronic problem
efficiently in the basis of the dimer states and allows properties such as the
BO potential to be studied as a function of the single coordinate~$R$. However,
such control on the atom positions is very difficult to achieve in
experiments~\cite{eiles}.


Here we study the high-$l$ trilobite states in a dense gas where the electron interacts with a large number of randomly positioned atoms. We show that highly polar, polyatomic
trilobite states not only exist in high density, but counterintuitively, their
dipole moment even increases with increasing density. We also show that a high
density provides a novel and general way to mix s-state character into the
trilobite states and thus make them accessible with conventional spectroscopy.

Our findings are explained by two key effects. Firstly, the recently discovered
mechanism of perturbation-induced quantum scarring~\cite{luukkoscars} explains
why perturbations from randomly arranged atoms around the Rydberg excitation
promote the formation of ordered eigenstates as a result of classical
symmetries in the unperturbed system. Secondly, inevitable random fluctuations
in the local density of the gas produce loose clusters that collectively
attract electron density. Combined, these effects cause surprisingly polar
``super-trilobite'' states to commonly form in a sufficiently dense random gas.


Atomic units are used throughout unless otherwise noted.

\section{Computational model}
To isolate the essential trilobite physics in a high density we use a model
that minimally generalizes the original dimer trilobite
calculations~\cite{trilobite} to more constituent atoms. The Hamiltonian of the
Rydberg-excited electron is~$H=H_0 + V$ where~$H_0$ contains the Coulomb
interaction with the Rydberg core and the quantum defects, and~$V = \sum_p
\Vs(\vec{R}_p)$ is the total perturbation from~$N$ atoms fixed in space at
locations~$\{\vec{R}_p\}$. In this we assume that the polarizations of atoms
are independent. Nearby atoms can, depending on the geometry, either increase
or suppress each others' polarization, but for the densities studied here this
should be a small effect compared to the direct polarization by the electron.

Because the range of the electron-atom interaction~$\Vs$ is very short compared to
the extent of the Rydberg electron, it can be modeled with the
Fermi pseudopotential~\cite{fermi}
\begin{equation}
  \label{eq:fermi}
  \langle\Psi_2|\Vs(\vec{R})|\Psi_1\rangle \coloneqq 2\uppi\as\Psi_2^*(\vec{R})\Psi_1(\vec{R}),
\end{equation}
where~$\as$ is the s-wave scattering length of the perturbing atom. The Fermi
model can be improved by adding terms corresponding to higher partial wave
scattering~\cite{omont}. The next term, corresponding to p-wave scattering,
depends on the gradients of the wave-functions at position~$\vec{R}$, and gives
rise to three ``butterfly states'' in the dimer case~\cite{hgs}. However, it is
the first term~\eqref{eq:fermi} that gives rise to the original trilobites. The
p-wave term also formally diverges at the location of shape resonances.

To model the energy-dependence of~$\as$ we use the modified effective range
theory~\cite{omalley} expression
\begin{equation}
  \label{eq:mert}
  \as(k) = \as(0) + \tfrac{\uppi}{3}\alpha k,
\end{equation}
where~$\alpha$ is the polarizability of the atom. The wave number~$k$ is set by
the semiclassical relation~$E = k^2/2 - R^{-1}$, where~$R$ is the distance to
the Rydberg core. As the atomic species we use $^{87}$Rb with~$\as(0) =
-16.1$ (spin triplet scattering) and $\alpha = 319.2$~\cite{bahrim}. Quantitatively more accurate
expressions for~$\as$ exist~\cite{khuskivadze}, but Eq.~\eqref{eq:mert}
provides a simple analytical formula that contains the key qualitative
behavior: atoms near the classical turning point ($k$ small) are most
attractive ($\as$ most negative), and the strength of the attraction decreases
at smaller~$R$.

The total perturbation~$V$ is not strong enough to mix states in different
degenerate hydrogenic manifolds for the densities and principal quantum numbers
considered here. For solving the eigenpairs of~$H$ for a given $n$ we therefore use a
basis of all hydrogen eigenstates whose energy, taking into account quantum
defects, is $E \in \left]E_{n-1}, E_n\right]$, where $E_n = -1/(2n^2)$. For the
Rb quantum defects we use values tabulated in Ref.~\onlinecite{khuskivadze}.

The positions $\{\vec{R}_p\}$ of the perturbing atoms are sampled within the classically allowed
radius $r_\text{c} = 2n^2$ from a Poisson point process with a
uniform density~$\rho$. Some atoms will fall too close to the Rydberg core for
Eq.~\eqref{eq:mert} to be accurate, but due to the overall diminishing of~$\Vs$
at small~$R$ these atoms do not contribute significantly.

\begin{figure}[tb]
  \centering
  \includegraphics[width=\linewidth]{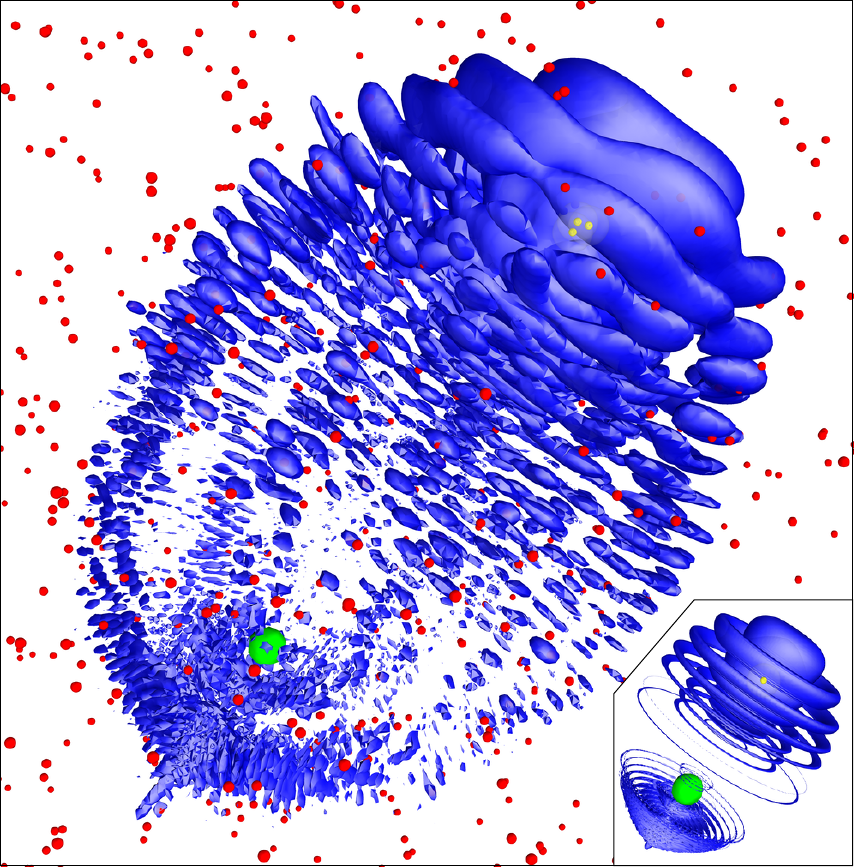}
  \caption{Electron density of a highly polar $n=50$ rubidium
	trilobite state at atom density $\rho =
	10^{16}~\si{\per\cubic\centi\metre}$. The electron density is visualized
	using a contour surface (in blue) that contains~42\% of the total
	probability mass. The small (red) spheres denote the locations of the
	perturbing atoms. For visibility, their radius is approximately twice the
	electron-atom scattering length. The electron density is localized around
	the highlighted cluster of three atoms (in yellow). The Rydberg core (in
	green) is at the origin, and the radius of the spherical cloud of atoms is
	the classically allowed radius~$r_\text{c} = 5000$. For comparison, the
	inset shows an equal visualization of the ordinary trilobite
	state~\cite{trilobite} with~$n=30$ and a single atom at~$R=1232$.
	Visualized with Mayavi~\cite{mayavi}.}
  \label{fig:state}
\end{figure}

\section{Trilobite states in high density}
The trilobite state is the lowest energy eigenstate of~$H$ that forms out of
the quasidegenerate high-$l$ states ($l\geq 3$ for Rb). For~$N=1$ this state
localizes around the single ground-state atom and acquires a large dipole
moment. Simple intuition suggests that, since the atoms are arranged randomly
with a uniform density, a large number of additional atoms simply smears out
the electron density into a featureless blob with zero dipole moment. Instead,
solving the eigenstates reveals that they are
often highly polar. At $n=50$ and $\rho = 10^{16}~\si{\per\cubic\centi\metre}$,
corresponding to over~$10^3$ atoms in the classically allowed volume,
84\% of trilobite states have a dipole moment greater than
$1~\text{kilodebye~(kD)}$. Figure~\ref{fig:state} shows an example of a
trilobite state  with dipole moment~$|\vec{d}| = 6~\text{kD}$ (2360~a.u.)
\footnote{See Supplemental Material at [URL will be inserted by publisher] for
a video of the trilobite state shown in Fig.~\ref{fig:state} and for results
with other principal quantum numbers~$n$.}. Far from a featureless blob, the
state resembles the dimer trilobite, except that the cylindrical
symmetry of the~$N=1$ case is broken into approximately elliptic
tracks of high probability density that orbit the Rydberg core.

To explain this oddly ordered structure it is beneficial to notice a connection
to a phenomenon recently observed in a related physical setting. Impurities perturbing a two-dimensional quantum dot produce ``scarred''
high-energy eigenstates, i.e., eigenstates that concentrate around paths of
classical periodic orbits (POs), even when the impurities are randomly
placed~\cite{luukkoscars}. This \emph{perturbation-induced (PI) scarring} is
similar in appearance to ordinary quantum scarring~\cite{scarsorig,
scarsreview}, but a key difference is that in PI scarring the corresponding
classical PO exists only in the \emph{unperturbed} system and the scars exist
only in the~\emph{perturbed} system -- perturbing the system with impurities
unmasks a hidden classical order.

The mechanism of PI scarring~\cite{luukkoscars} requires two general
ingredients, both of which are present in the current system: the unperturbed
system needs to be separable, and the individual perturbations need to have a
short spatial range.

If the unperturbed system is separable, applying the semiclassical
Bohr--Sommerfeld quantization formula shows that at high energies its
eigenstates are bunched into quasidegenerate sets. Each such ``resonant set''
corresponds to a family of classical POs, and this connection ensures that some
linear combinations of the resonant states are scarred by corresponding POs.

A moderate perturbation produces eigenstates that are linear combinations of a
single resonant set. By the variational theorem, the eigenstates corresponding
to extremal eigenvalues extremize the Hamiltonian. Because the resonant states
are nearly degenerate, this essentially extremizes the perturbation. In this
extremization the scarred states have an advantage if the perturbations have a
short range: scarred states can very effectively maximize (minimize) the
perturbation by selecting paths that hit atypically many (few) perturbations.
As a result, the extremal eigenstates arising from each resonant set often
contain scars of the corresponding PO.

In the present system the separable unperturbed system is a hydrogen-like
Rydberg atom, and the short-range electron-atom interactions create the
localized perturbations. The resonant sets are the constant-$n$ manifolds
\footnote{The Bohr--Sommerfeld formula happens to be exact for hydrogen, making
the resonant sets perfectly degenerate in the absence of quantum defects.}, the
corresponding POs are Kepler orbits with a fixed energy, and the maximally
scarred linear combinations are the stationary elliptic states~\cite{gay}.

The trilobite state essentially minimizes~$V$ within the set of quasidegenerate
high-$l$ states, i.e., it maximizes the electron density on the perturbing
atoms. However, since it is confined within the quasidegenerate set, it has
only limited capability to do so. The eigenstates scarred by Kepler orbits --
guaranteed by semiclassics to exist -- often represent the best available
solution. Since a single Kepler orbit is confined to a plane, a single scar can
only coincide with a small fraction of the perturbing atoms. Therefore it is
unlikely that eigenstates localize perfectly on a single orbit, as in
Ref.~\onlinecite{luukkoscars}, but instead a scarred state likely contains
scars of several orbits. The connection between Kepler orbits and the dimer
trilobite state was also studied in Ref.~\onlinecite{granger}.

If several atoms perturb the system it is also important to consider their
relative weights in the minimization of~$V$. This weight is essentially given
by the average value of $\Vs(\vec{R}_p)$, i.e., the dimer BO potential at
interatomic distance~$R_p$. Atoms at positions corresponding to a deep dimer
potential give a larger contribution to the total~$V$ and are thus favored,
improving the localization.

A final and crucial point is that while the overall mean density of atoms
is uniform, the individual random snapshots of~$\{\vec{R}_p\}$ are not.
Random fluctuations in the local density produce clusters where a few atoms
happen to reside close to each other. At large~$R$ the local wavelength of the
electron is large, and even a relatively loose cluster of atoms can fit completely
inside one half-wave. Such a cluster then acts as a single ``superperturber''
that can completely dominate in the minimization of~$V$. As an example, the
state in Fig.~\ref{fig:state} is strongly localized around a cluster of three
atoms located within 130~a.u.\ of each other. Because the available volume
($\propto n^6$) is large, the existence of such clusters has a surprisingly
high probability. This is essentially a spatial analogue of the famous
birthday paradox~\cite{birthday}.

It is important to note that the clusters can be fairly sparse. At~$n=50$
clusters with atomic separations between 100 and 300 a.u.\ seem to be common in
high dipole moment trilobite states, and even larger clusters are possible for
higher~$n$.

The dominating nature of the clusters simplifies estimating the stability of
the polar trilobite configuration, as it is sufficient to study only the
few-body problem of the cluster. However, this is still substantially more
difficult than in the dimer case. Adding a p-wave scattering term to~$H$ also
likely produces a separate class of ``butterfly states'', just as in the dimer
case~\cite{hgs}. The states separate into s-wave dominated trilobites and
p-wave dominated butterflies because the same state cannot simultaneously
maximize both the wave function and its gradient at the position of the
cluster.

The clusters complete the explanation for highly polar trilobite states at high
density. There is a high probability that a cluster of atoms occurs somewhere
in the large volume of space where the dimer potential is deep, and this
cluster then dominates in the minimization of~$V$. The states within the
quasidegenerate high-$l$ manifold that most efficiently maximize the electron
density on the cluster are scars of Kepler orbits whose apoapsis, where the
electron density is higher, occurs near the cluster. Atoms not in the cluster
have only a minor role in the formation of the state, but they do break the
cylindrical symmetry and give the trilobite a skeletal appearance in which
individual Kepler orbit PI scars can be resolved, as in Fig~\ref{fig:state}.

\begin{figure}[tb]
  \centering
  \includegraphics{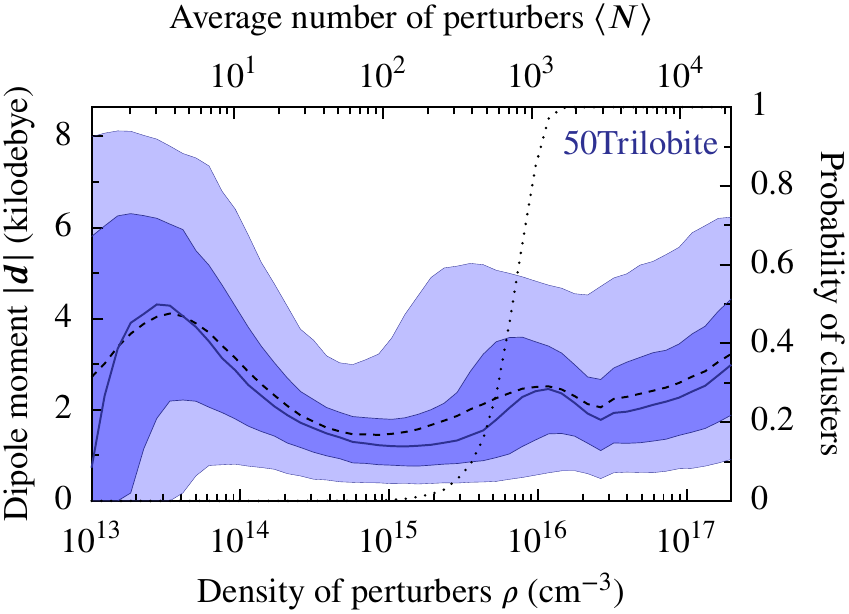}
  \caption{Probability distribution of the dipole moment of the $n=50$
	trilobite state as a function of the atom density~$\rho$. The solid line
	shows the median value and the edges of the two shaded regions denote the 5th,
	25th, 75th, and 95th percentiles. The dashed line denotes the mean value.
	The dotted line shows the probability that a snapshot of~$\{\vec{R}_p\}$
	contains at least one cluster where three atoms are located within 300~a.u.\ of each other.
	The values are estimated from an ensemble of $10^4$ snapshots of the
	atom locations.}
\label{fig:dm}
\end{figure}

\section{The dipole moment}
A key property that makes the dimer trilobites interesting is their
extraordinarily large dipole moment. An important consequence of the previously
described mechanism is that the trilobite states can retain this property at
high densities. Computing the dipole moment also provides a way to quantify how
common the polar trilobite states are for a given density~$\rho$.

Figure~\ref{fig:dm} shows the probability distribution of the dipole
moment~$|\vec{d}|$ as a function of~$\rho$ for the $n=50$ trilobite state. As
the average number of perturbing atoms~$\langle N\rangle$ increases, the
competition between atoms causes the dipole moment to initially decrease.
However, as clusters of atoms become more likely the dipole moment starts to
increase as a function of~$\rho$. As non-intuitive as it is, adding atoms
randomly and uniformly in space makes the trilobite states \emph{more} polar.
Other principal quantum numbers~$n$ give similar results~\cite{Note1}.
Fig~\ref{fig:dm} also shows how the probability of a sufficiently small cluster
of three atoms grows as a function of~$\rho$. The existence of a cluster
\emph{somewhere} in the classically allowed region is almost certain at $\rho =
10^{16}~\si{\per\cubic\centi\metre}$, even though the probability of three
atoms existing in a \emph{given}, cluster-sized volume is still very small
($10^{16}~\si{\per\cubic\centi\metre} = \num{1.5e-9}\ \text{a.u.}$).

\begin{figure}[tb]
  \centering
  \includegraphics{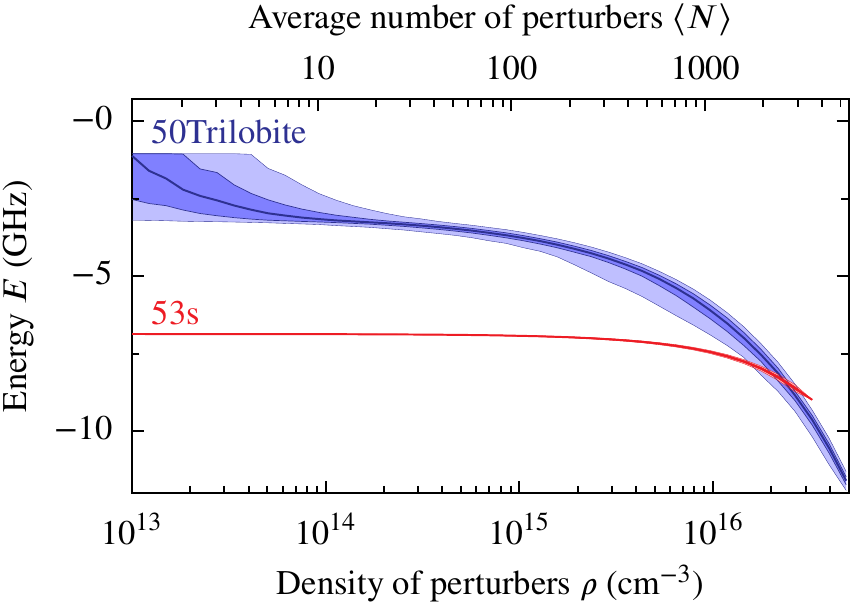}
  \caption{Probability distribution of the energy of the $n=50$ trilobite
	state and the $53s$ state as a function of the atom density~$\rho$. The
	zero of energy is chosen at the degenerate $n=50$ hydrogen manifold. The
	probability distribution is visualized through its mean and percentile
  curves and estimated from an ensemble as in Fig.~\ref{fig:dm}.}
  \label{fig:energy}
\end{figure}

\begin{figure}[tb]
  \centering
  \includegraphics{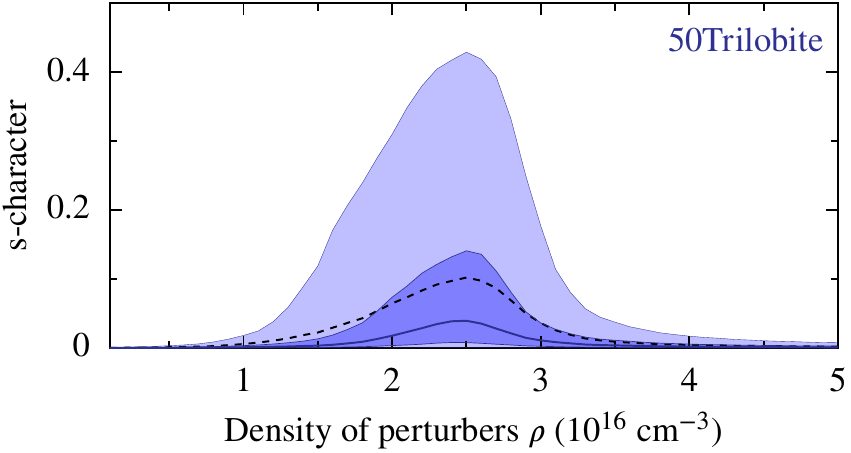}
  \caption{Distribution of the magnitude of $l=0$ character in the $n=50$
	trilobite state, i.e., the squared overlap between the unperturbed 53s
	state and the trilobite state, as a function of the atom density~$\rho$.
	The probability distribution is visualized through its mean and percentile
  curves and estimated from an ensemble as in Fig.~\ref{fig:dm}.}
  \label{fig:s}
\end{figure}

\section{Energy shift and mixing with the s-state}
Comparing the behavior of high-$l$ states and s-states in a high density uncovers an important and
experimentally relevant detail of the high-density trilobites.
Figure~\ref{fig:energy} shows the energy of the~$n=50$ trilobite state and the
energetically neighboring~53s state as a function of~$\rho$. Interaction with
the perturbing atoms shifts both states to lower energies and, as predicted by
Fermi~\cite{fermi}, the s-state energy shift depends linearly
on~$\rho$~\cite{liebisch}. However, as already seen in the previous results,
the quasidegenerate manifold of high-$l$ states has much more freedom to optimize
the electron density around variations in the local density. As a result the
trilobite state shifts down in energy faster than the s-state as~$\rho$ is
increased.

As seen in Fig.~\ref{fig:energy}, the $n=50$ trilobite crosses the~53s state
near $\rho=\SI{2e16}{\per\cubic\centi\metre}$, creating hybrid states that
contain some s-character. This is shown in Fig.~\ref{fig:s} in more detail.
Mixing with non-polar s-states also causes a dip in the dipole moment at this
density in Fig.~\ref{fig:dm}. As mentioned, mixing s-character into the
trilobites is a highly sought-for property~\cite{booth,eilesgreene,kleinbach}
because without low-$l$ character the photoassociation of trilobite molecules
is hindered by selection rules. In previous works it has been achieved by
choosing a system with special resonances, but here it arises as a natural
consequence of the variational flexibility of the high-$l$ states.

\section{Outlook}
An atom density of $\rho \approx 10^{16}~\si{\per\cubic\centi\metre}$ at an
ultracold temperature is, to our knowledge, approximately an order of magnitude
higher than what is realized in current experiments. With higher~$n$ the local
electron wavelength at large~$R$ and the available volume are larger, and thus
the clustering regime is reached at lower densities~\cite{Note1}. On the other hand,
$n\approx50$ has the advantage that the clustering regime and the s-mixing
regime occur at similar densities. As a high~$\rho$ is only needed to make
random clusters more probable in a homogeneous density, one possible option is
to specifically increase the local density at some points.

To summarize, we have demonstrated that highly polar trilobite states persist
and even thrive in a dense gas. This counterintuitive result emerges from a
combination of perturbation-induced quantum scarring and randomly occurring
atom clusters in the gas. Moreover, at a certain $n$-dependent density of the
gas the states gain some s-character and thus become accessible with
conventional photoassociation, overcoming a common obstacle with conventional
dimer trilobites.

\begin{acknowledgments}
  We thank B.~Drury for useful discussions.
\end{acknowledgments}

\bibliography{triloscars-paper}

\iftoggle{includesupplementary}{%
\appendix
\clearpage
\begin{center}
  \makeatletter
  \large\textbf{\@title}\\[0.5em]
  \normalsize\textbf{Supplementary material}
  \makeatother
\end{center}

\section{Other principal quantum numbers} The main article describes trilobite
states forming from the quasidegenerate manifold of states with principal
quantum number~$n=50$. For completeness,
Figures~\ref{fig:dm_n40}--\ref{fig:energy_n90} show the probability
distributions of the trilobite state energy and dipole moment for other values
of~$n$. All cases show a similar qualitative behavior: highly polar trilobite
states start to localize on clusters at some sufficiently high atom density,
and at some density the states cross and mix with a s-state. For larger~$n$,
larger clusters can fit inside a single half-wavelength of the electron, and
consequently highly polar trilobite states appear at lower atom densities. The
calculations for~$n=70$ and $n=90$ use a smaller ensemble of 1000 snapshots
of~$\{\vec{R}_p\}$.

For higher~$n$ the cluster-induced trilobite states recover a smaller fraction
of the dipole moment of the dimer case ($\langle P\rangle~\approx~1$). This is
because as~$n$ increases, the global minimum of the dimer Born--Oppenheimer
potential occurs at a smaller fraction of $r_\text{c} = 2n^2$. The radii near
this minimum is where the cluster can dominate over other perturbing atoms. It is
important to note that these radii are also where \emph{stable} dimer molecules
are found, i.e., the cluster-induced trilobite states have similar dipole
moments as the dimer trilobites in a stable configuration, even at higher~$n$.

\begin{figure*}[pb]
  \begin{minipage}{0.48\linewidth}
  \centering
  \includegraphics{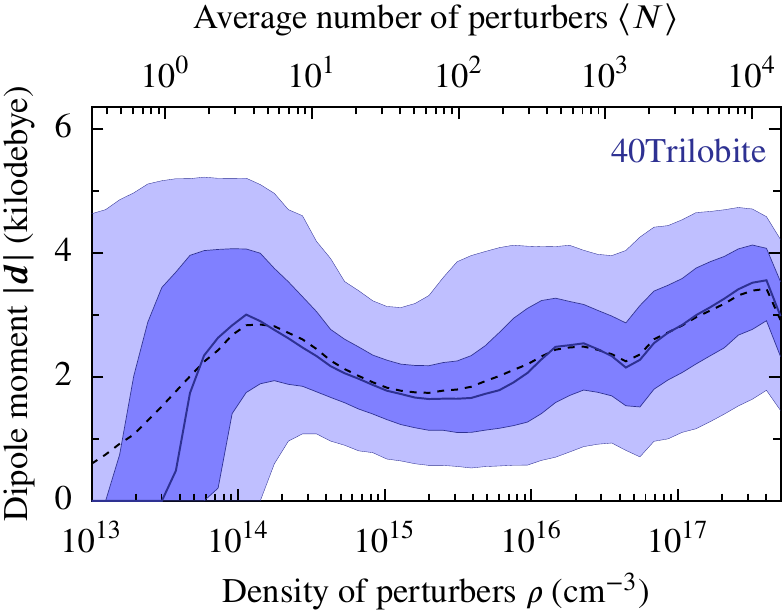}
  \caption{Probability distribution of the dipole moment, as Fig.~2 but with $n=40$.}
  \label{fig:dm_n40}
  \end{minipage}\hfill
  \begin{minipage}{0.48\linewidth}
  \centering
  \includegraphics{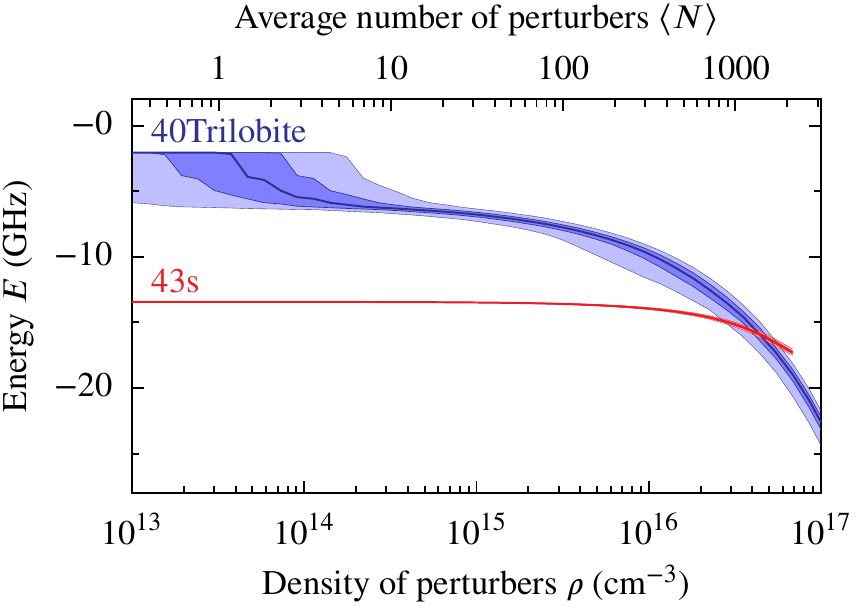}
  \caption{Probability distribution of the energy, as Fig.~3 but with $n=40$.}
  \label{fig:energy_n40}
  \end{minipage}
\end{figure*}

\begin{figure*}[tpb]
  \begin{minipage}{0.48\linewidth}
  \centering
  \includegraphics{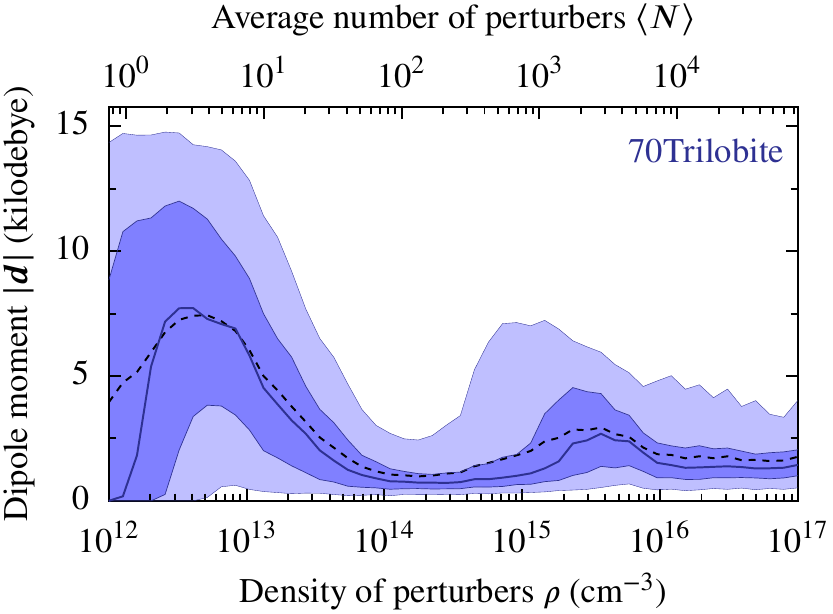}
  \caption{Probability distribution of the dipole moment, as Fig.~2 but with $n=70$.}
  \label{fig:dm_n70}
  \end{minipage}\hfill
  \begin{minipage}{0.48\linewidth}
  \centering
  \includegraphics{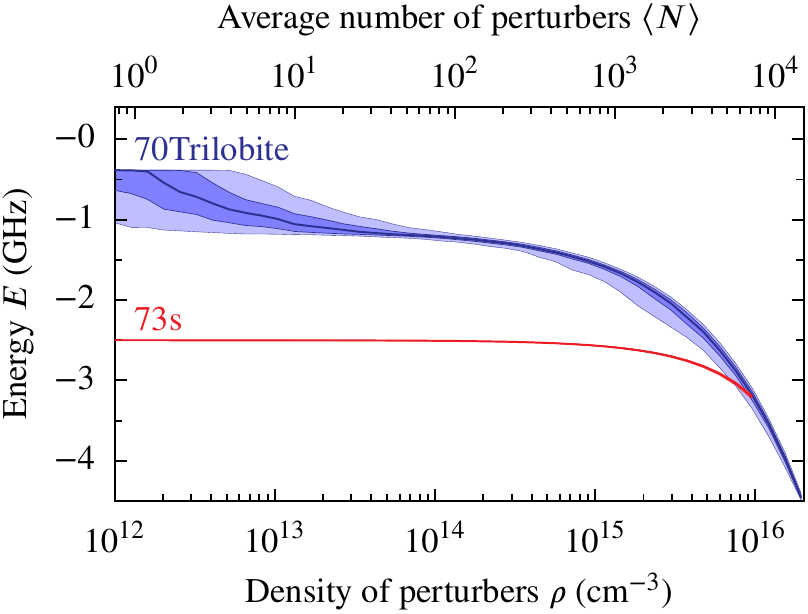}
  \caption{Probability distribution of the energy, as Fig.~3 but with $n=70$.}
  \label{fig:energy_n70}
  \end{minipage}
\end{figure*}

\begin{figure*}[tpb]
  \begin{minipage}{0.48\linewidth}
  \centering
  \includegraphics{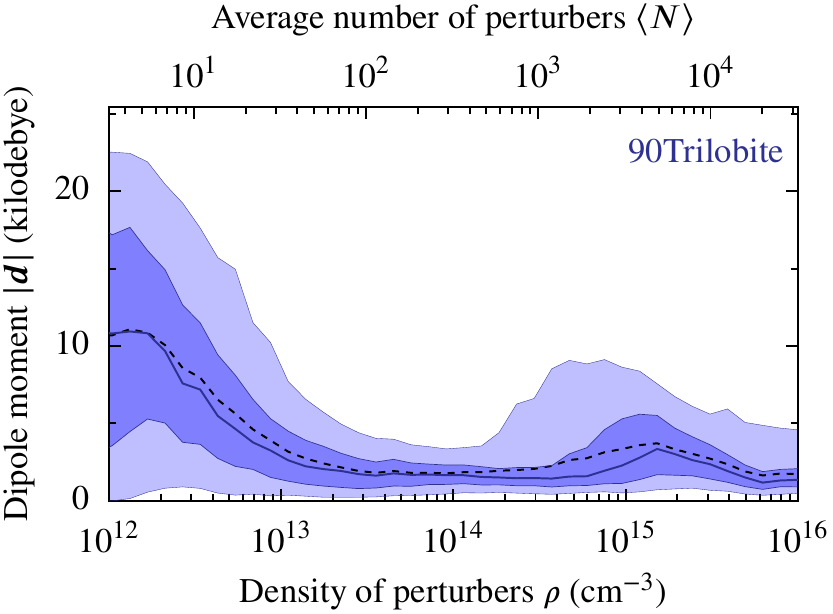}
  \caption{Probability distribution of the dipole moment, as Fig.~2 but with $n=90$.}
  \label{fig:dm_n90}
  \end{minipage}\hfill
  \begin{minipage}{0.48\linewidth}
  \centering
  \includegraphics{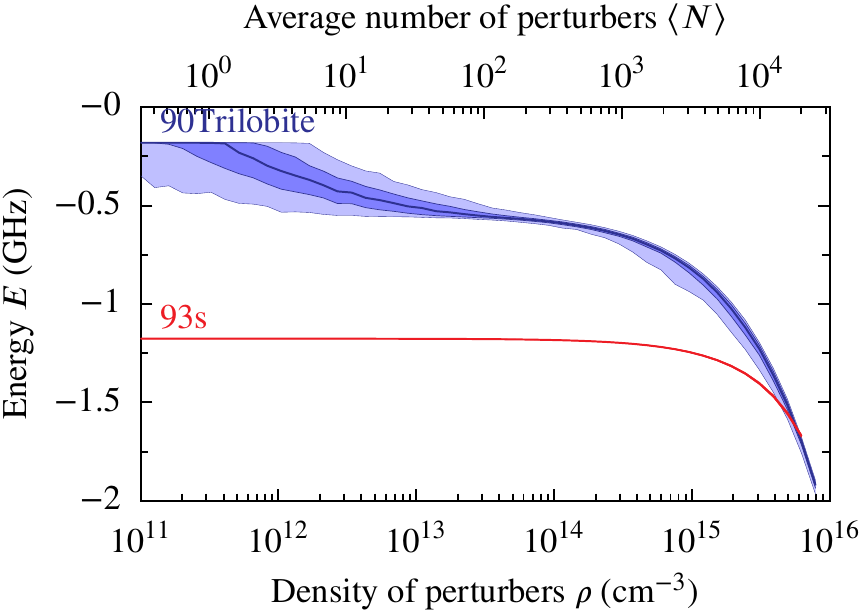}
  \caption{Probability distribution of the energy, as Fig.~3 but with $n=90$.}
  \label{fig:energy_n90}
  \end{minipage}
\end{figure*}

\section{Importance of clusters}
As discussed in the main article, in a high atom density the trilobite states
localize on randomly occurring clusters of atoms. Figure~\ref{fig:mind}
illustrates the importance of clusters by showing the probability distribution
of the trilobite state dipole moment in a calculation where most clusters are
excluded by enforcing a minimal distance between atoms. Without clusters there
is no increase in dipole moment at high densities.

\begin{figure}[tpb]
  \centering
  \includegraphics{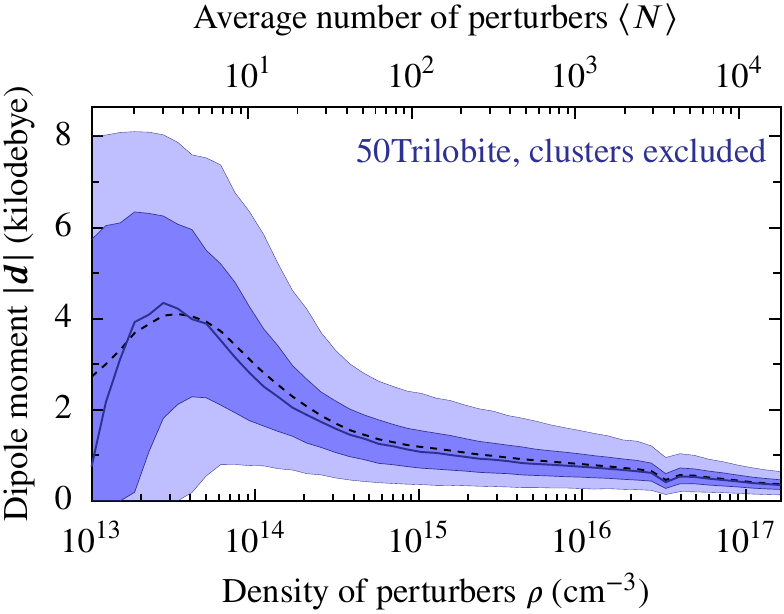}
  \caption{Probability distribution of the dipole moment, as Fig.~2 but
  excluding atom clustering by enforcing a minimal distance of 300~a.u.\
between atoms.}
  \label{fig:mind}
\end{figure}

}{}

\end{document}